\documentclass[letterarticle, 10 pt, conference]{ieeeconf}

\newcommand{\booktitle}{Reachability Analysis for FollowerStopper: \\ Safety Analysis and Experimental Results}

\usepackage{multirow}
\usepackage{amsmath}
\usepackage{booktabs}
\usepackage[margins]{trackchanges}
\usepackage{rahulbhadani}
\usepackage{mdframed}
\graphicspath{ {figures/} }

\usepackage{caption}
\usepackage{subcaption}
\makeatletter
\let\NAT@parse\undefined
\makeatother
\usepackage[bookmarks,bookmarksnumbered,
pdfborder={0 0 0},
    linktocpage,
    colorlinks=true, citecolor=red,urlcolor=blue,bookmarks=true,hypertexnames=true]{hyperref}
\author{Fang-Chieh Chou, Marsalis Gibson, Rahul Bhadani$^{\ddagger}$, Alexandre M. Bayen and Jonathan Sprinkle$^{\dagger}$ 
\thanks{Fang-Chieh Chou is affiliated with the Department of Mechanical Engineering, University of California, Berkeley, CA, USA.  Email: \texttt{fcchou@berkeley.edu}.
Marsalis Gibson is affiliated with the Department of Electrical Engineering, University of California, Berkeley, CA, USA. Email: \texttt{mtgibson@berkeley.edu}
Alexandre M. Bayen is affiliated with the Department of Electrical Engineering, University of California, Berkeley, CA, USA. Email: \texttt{bayen@berkeley.edu} 
$^{\ddagger}$Rahul Bhadani is affiliated with the Department of Electrical and Computer Engineering, The University of Arizona, USA. Email: \texttt{rahulbhadani@email.arizona.edu}}%
\thanks{$^{\dagger}$Jonathan Sprinkle is affiliated with the Department of Electrical and Computer Engineering, The University of Arizona, USA. Email: \texttt{sprinkjm@email.arizona.edu}}}

\title{\LARGE \bf \booktitle}

\IEEEoverridecommandlockouts
\newcommand{\disturbanceConstraintSet}{\ensuremath{\mathcal{D}}}
\newcommand{\stateConstraintSet}{\ensuremath{\mathcal{Z}}}
\newcommand{\inputConstraintSet}{\ensuremath{\mathcal{U}}}

\newcommand{\state}{z}
\newcommand{\sysinput}{u}

\newcommand{\disturbance}{d}

\begin{document}

\maketitle

\thispagestyle{empty}
\pagestyle{empty}

\begin{abstract}
Motivated by earlier work and the developer of a new algorithm, the FollowerStopper, this article uses reachability analysis to verify the safety of the FollowerStopper algorithm, which is a controller designed for dampening stop-and-go traffic waves. With more than 1100 miles of driving data collected by our physical platform, we validate our analysis results by comparing it to human driving behaviors. The FollowerStopper controller has been demonstrated to dampen stop-and-go traffic waves at low speed, but previous analysis on its relative safety has been limited to upper and lower bounds of acceleration. To expand upon previous analysis, reachability analysis is used to investigate the safety at the speeds it was originally tested and also at higher speeds. Two formulations of safety analysis with different criteria are shown: distance-based and time headway-based. The FollowerStopper is considered safe with distance-based criterion. However, simulation results demonstrate that the FollowerStopper is not representative of human drivers - it follows too closely behind vehicles, specifically at a distance human would deem as unsafe. On the other hand, under the time headway-based safety analysis, the FollowerStopper is not considered safe anymore. A modified FollowerStopper is proposed to satisfy time-based safety criterion. Simulation results of the proposed FollowerStopper shows that its response represents human driver behavior better.




\end{abstract}

\section{Introduction}

The field operational test \cite{stern2018dissipation} demonstrated the ability of a single autonomous vehicle to dampen traffic waves that emerged as part of human-driven traffic on a specially constructed ring road. One controller used in that experiment, the FollowerStopper \cite{Rahul2018}, used recorded driver data from the same set of drivers to establish bounds on the estimated maximum decelerations that might be observed in a ring road drive. Additional work has further demonstrated that vehicle controllers can learn to dampen traffic waves on road networks, through simulated drives using intelligent driver models to simulate traffic \cite{wu2019flow}. 

There are several limitations of these previous results. The FollowerStopper has never been tested in a physical experiment at speeds more than $8.0\;m/s$, and its design might have safety flaws that are as yet unexplored. Second, the data used to design the FollowerStopper came from ring road drives of the same ring radius, and not from naturalistic human driving, leading to a potential mismatch.

This article explores how to transition from the ring road to the physical highway for the FollowerStopper by addressing these two limitations. First, we apply reachability analysis to the FollowerStopper controller (as described in \cite{stern2018dissipation}), to determine whether it is safe at velocities that are more typical of transportation networks. Secondly, we use data gathered from over 180 drives totally over $1100$ miles by the same driver at speeds varying between $[0,30]m/s$, to explore how the human-driven data can be compared with the existing FollowerStopper's design.

\begin{mdframed}
The contributions of the work are (i) a demonstration that the existing FollowerStopper is not safe enough to deploy, and (ii) a new design for the FollowerStopper, with reachability analysis and simulation to demonstrate its safety.
\end{mdframed}
The reachability analysis performed on the existing design demonstrates why additional safety criterion must be added in order to be safe at speeds larger than those seen in the ring road experiments. The new design uses driver headways at various speeds observed from naturalistic driving data, to establish a velocity-varying constraint on the controller's adaptation regions. The reachability analysis of the new controller shows that the new design more closely encapsulates the naturalistic driving data. The result is a controller that, when simulated, exhibits following behaviors more typical of human drivers when compared to the original design.

\subsection*{Organization}
In section \ref{sec:background}, we provide brief reviews of relevant works. We formulate the safety analysis for FollowerStopper in section \ref{sec:problemFormulation}. The distance-based safe set of FollowerStopper is shown in section \ref{sec:safetyAnalysis4FollowerStopper}, along with comparing against driving data. Time headway-based safety analysis, which factors in the absolute speed of the subject vehicle is shown in section \ref{sec:safetSetWithTimeHeadway}. To meet time headway-based safety, FollowerStopper is modified and verified with time headway safety analysis in section \ref{sec:Modify_FollowerStopper}. Section \ref{sec:conclusions} concludes this article.


\section{Background} \label{sec:background}

\subsection{Prior work in the safety of vehicle-following control}

Vehicle-following control is a class of \textit{advanced driver assistance systems} that automatically control the relative space and/or relative speed between the subject vehicle and the leading vehicle. \textit{Adaptive cruise control} (ACC) \cite{Xiao2010_reviewACC} is a type of most popular vehicle-following control, which controls the longitudinal motion of the subject vehicle based on measurements of relative distance and relative speed to the leading vehicle.  

For ACC consisted of spacing policy, which governs the desired gap at a steady state, the spacing policy is critical to safety. By designing the spacing policy appropriately, safety can be enhanced; for example, constant safety factor spacing policy picks the spacing based on the minimum safe stopping distance so that subject vehicle can stop safely if the leading vehicle 
make a hard deceleration \cite{Wu2020_ACCPolicies}.  
However, not every car following controller has explicitly-defined spacing policy. For linear car-following models, safety can be guaranteed by verifying impulse response of the model \cite{Lunze2019}. Ideally, if we can predict the trajectory of the leading vehicle in the future, we can verify the car-following safety by checking if there is any collision between the subject vehicle and the leading in the future. Usually, a few representative trajectories of the leading vehicle and the following vehicle are used to assess safety  \cite{MAGDICI2017} \cite{Ligthart2018} \cite{Willigen2015} \cite{Willigen2011}. 

In this article, we use reachability analysis to evaluate safety with regard to all possible motions of the leading vehicle and the following vehicle. We apply this method to verify safety of the FollowerStopper.

\subsection{Review of the FollowerStopper controller}
\label{sec:FollowerStopper}

\begin{figure}[htpb]
    \centering
    \vspace{+0.5em}
    \includegraphics[width=0.45\textwidth]{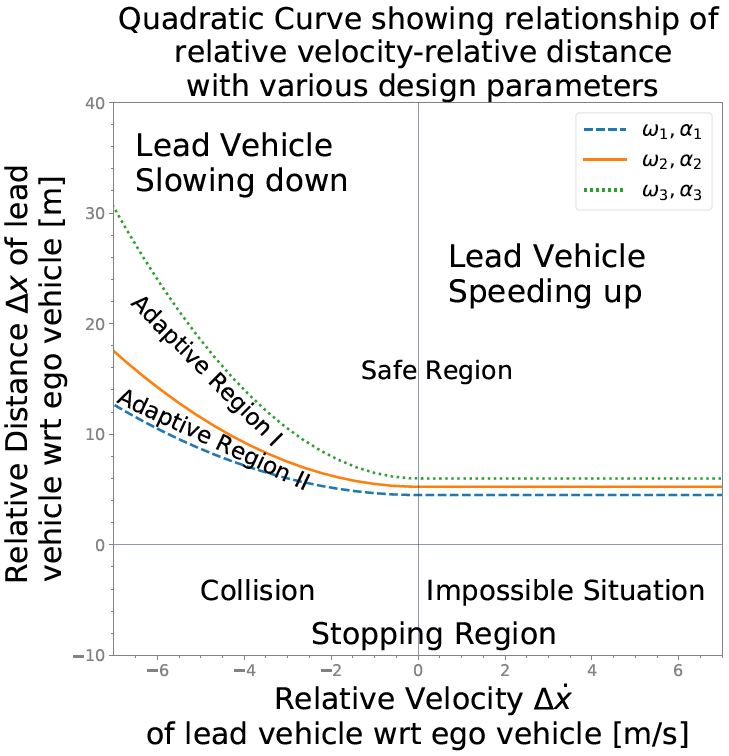}
    \caption{Quadratic curve with various switching region of FollowerStopper on relative distance-relative phase space.}
    \vspace{-0.5em}
    \label{fig:quadratic_curve}
\end{figure}

The FollowerStopper\cite{Rahul2018}\cite{bhadani2019real} is a car following controller designed for low-speed operations ($v<10\text{m/s}$) that computes the desired speed of the subject vehicle $v_{cmd}(t) \in \mathbf{R}$, given the relative distance $x_{rel}(t) \in \mathbf{R}$ and the relative speed $v_{rel}(t) \in \mathbf{R}$ between the leading vehicle and the subject vehicle, at time $t \in [0,\infty)$. For a given relative speed $\dot{\Delta x} \in \mathbf{R}$, the state space of relative distance $\Delta x \in \mathbf{R}$ is divided into four zones separated by 
$\bar{x}_1$, $\bar{x}_2$, and $\bar{x}_3$, which are defined as:
\begin{align}
    \bar{x}_j = \omega_j + \frac{1}{2\alpha_j} {v_{rel}^*}^2, \quad j = 1,2,3.  \label{eq:FS_segmentation}
\end{align}
where $\omega_j \in \mathbf{R}$, $\alpha_j \in \mathbf{R}$ are design parameters, and $v_{rel}^* = \min\{\dot{\Delta x},0\}$, 
Different zones in the quadratic band allows ego vehicle to transition from one mode to another smoothly without discomfort to the driver or passenger. This idea is illustrated in Figure~\ref{fig:quadratic_curve}.


\begin{figure}[tpb]
    \centering
    \includegraphics[trim={0.6cm 0.0cm 3.0cm 0.0cm},clip, width=1.0\linewidth]{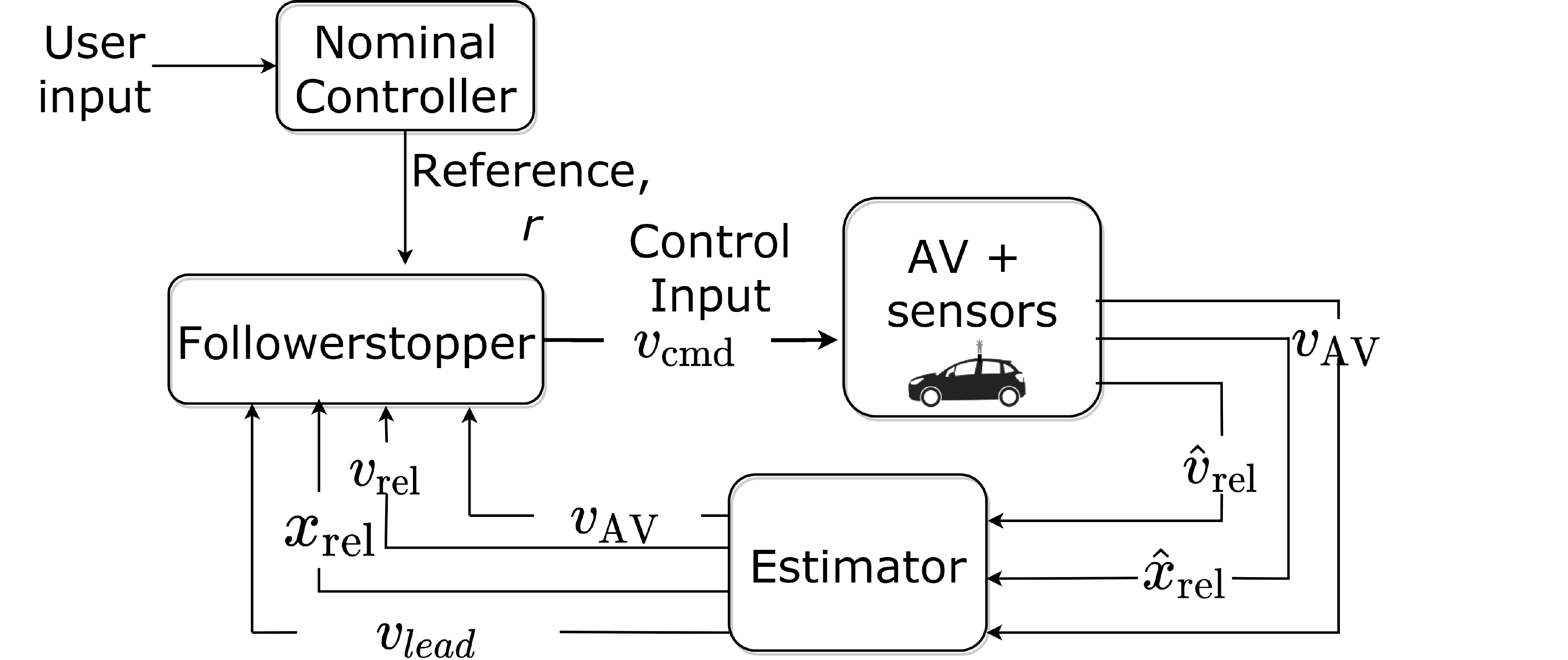}
    \vspace*{8pt}
    \caption{A schematic diagram of the FollowerStopper controller and other components used for vehicle control.}
    \vspace{-1.5em}
    \label{fig:controllermodel}
\end{figure}

Depending on which zone is valid for the current relative distance, the desired speed is computed as:
{\small
\begin{align}
    v_{cmd} (\Delta x,\dot{\Delta x}) = \begin{cases}
    0,  &\textrm{if  } \Delta x \leq \bar{x}_1\\
    v \cfrac{\Delta x-\bar{x}_1}{\bar{x}_2-\bar{x}_1}, &\textrm{if  }  \bar{x}_1 < \Delta x \leq \bar{x}_2\\
    v+(r-v)\cfrac{\Delta x-\bar{x}_2}{\bar{x}_3-\bar{x}_2}, &\textrm{if  } \bar{x}_2 < \Delta x \leq \bar{x}_3\\
    r, &\textrm{otherwise}
    \end{cases},
\end{align}}where $v= \min\{ \max\{v_{lead},0\},r \}$; $r$ is $30\text{m/s}$. 
Typically, vehicle speed is not exactly equal to  the command speed $v_{cmd}(t)$ because of the nature of the vehicle dynamics. We consider vehicle dynamics (section \ref{sec:vehicleDynamicsModelA}) in reachability analysis to capture real response of the vehicle. A schematic of the FollowerStopper model is given in Figure~\ref{fig:controllermodel}.

\section{Problem Formulation}
\label{sec:problemFormulation}

The work described in the rest of this article seeks to provide a more formal analysis to the design properties of the FollowerStopper: 
\begin{itemize}
    \item Are collisions avoided within maximum range of behavior uncertainty and velocity estimates?
    \item Are the deceleration, acceleration, and velocity profiles generated from the FollowerStopper, and the gaps followed by the FollowerStopper representative of human drivers?
\end{itemize}


We conduct a formal analysis of the safety of FollowerStopper, where safety is considered as capabilities of avoiding a set of undesirable states, particularly, undesirable states of the relative distance, the relative speed, and the subject vehicle speed. A safe set is a set of states from which avoidance of undesirable states can be guaranteed, which can be synthesized by using a Hamiltonian-Jacobi PDE. 

When formulating the reachability problem, there are a few challenges, such as estimates for the uncertainty and system parameters, and the model mismatch. In any formal analysis, there will always be a difference between the models used in the analysis and the true system's behavior in the real-world. We mitigate this difference by incorporating experimental driving data sets in the reachability formulation and for further analysis of the system. 


\subsection{Background}
Safety analysis of a controller may be formulated as a reachability problem, which can be written as a terminal cost differential game, where the reachable set is characterized as a sub-level set of a value function. The problem can be reduced to solving the time-dependent Hamilton-Jacobi partial differential equation (HJ PDE) \cite{mitchell2005time}, which in turn can be solved with the Level Set toolbox \cite{mitchell2007ToolBoxManual}.



\subsubsection{Reachable set}

Consider the system model with state $\state \in \stateConstraintSet \subset \mathbf{R}^n$, and input $\sysinput \in \inputConstraintSet \subset \mathbf{R}^{n_u}$; the model is subject to disturbance $\disturbance \in \disturbanceConstraintSet \subset \mathbf{R}^{n_d}$:
\begin{align}
    \dot{\state} = f(\state,\sysinput,\disturbance),
\end{align}
where $f:\stateConstraintSet \times \inputConstraintSet \times \disturbanceConstraintSet \rightarrow \mathbf{R}^n$. Let
$\xi_{\state_0}^{\bar{\sysinput}(\cdot),\bar{\disturbance}(\cdot)}: [0,\infty) \rightarrow \mathbf{R}^n$ denotes a trajectory of the system given an initial condition 
$z_0 \in \mathbf{R}^n$ with a input $\bar{\sysinput}(\cdot)$ and a disturbance $\bar{\disturbance}(\cdot)$, where $\bar{u}(\cdot)$ and $\bar{\disturbance}(\cdot)$ are drawn from sets of measurable functions: $\bar{\sysinput}(\cdot) \in \mathbb{U} = \{\mathbf{u}:[0,\infty) \rightarrow \inputConstraintSet~|~ \mathbf{u} \textrm{ is measurable}\}$, $\bar{\disturbance}(\cdot) \in \mathbb{D} = \{\mathbf{d}:[0,\infty) \rightarrow \disturbanceConstraintSet~|~ \mathbf{d} \textrm{ is measurable}\}$. 
For a given target set $\Omega \subseteq \stateConstraintSet$, a reachable set $\mathcal{R}$ is the set of state from which there exists some disturbance $\bar{\disturbance}$ such that the trajectory of the state will land into the target set for all possible control action $\bar{\sysinput}$. Formally, $\mathcal{R}$ is defined as follows:
\begin{equation}
    \begin{aligned}
        \mathcal{R} = \{z_0 \in \mathcal{Z} ~|~ \forall & \bar{u}(\cdot) \in \mathbb{U}, \exists \bar{d}(\cdot) \in \mathbb{D} \\
     \textrm{~s.t.~} & \xi_{\state_0}^{\bar{\sysinput}(\cdot),\bar{\disturbance}(\cdot)}(t) \in \Omega ~, \exists t \in [0,\infty)\}.
    \end{aligned}
\end{equation}


\subsubsection{Reachability analysis as a differential game}
Let the target set $\Omega$ be characterized by a value function $l(z):\mathbf{R}^n \rightarrow \mathbf{R}$, where values are negative for states inside the target set, positive for states outside the target set and zero on the boundary:
$\Omega = \{z \in \mathcal{Z} ~|~ l(z) \leq 0, ~l:\mathbf{R}^n \rightarrow \mathbf{R}\}.$
For a given $z_0$, $\bar{u}(\cdot)$, and $\bar{d}(\cdot)$ a value function $\mathcal{V}$ defined as follows can be used to determine whether trajectory have ever visited the target set:
\begin{align}
        \mathcal{V}(z_0,\bar{u}(\cdot),\bar{d}(\cdot)) = \inf_{t \geq 0} l(\xi_{z_0}^{\bar{u}(\cdot),\bar{d}(\cdot)}(t)).
\end{align}
The value function captures the lowest value that the trajectory has ever achieved, so that the worst case could be captured. The value would be positive, if the trajectory has never visited the target set $\Omega$; otherwise, it would mean the trajectory has visited the target set at least once. 

The reachability set can be considered as an optimal solution of a differential game of two players, where Player I with the control $\bar{u}(\cdot)$ is trying to drive the state $\xi_{z_0}^{\bar{u}(\cdot),\bar{d}(\cdot)}(\cdot)$ stay away from the target set $\Omega$ and Player II with the control of $\bar{d}(\cdot)$ is trying to drive the state inside the target set. Player I would win the game if the trajectory of the state has never visited the target set, and Player II would win the game if the trajectory has ever been to the target set. In other words, for a given $z_0 \in \mathbf{R}^n$, Player I would win the game if the outcome of the value function $\mathcal{V}$ is positive, and Player II would win the game if the outcome is negative. Therefore, the optimal control of Player I should try to maximize $\mathcal{V}$, while the optimal control of Player II should try to minimize $\mathcal{V}$. In the sense that the reachability set we are interested in is the worst case for Player I, we are favoring Player II, assuming Player II has information of Player I at every instance and can respond accordingly, but Player II does not know the future action of Player I ahead of time. We assume Player II uses non-anticipative strategies \cite{fisac2018general}\cite{LCEVANS1984} $\Delta$:
\begin{equation}
    \begin{aligned}
           &\Delta  = \{\delta : \mathbb{U}\rightarrow \mathbb{D} ~ | ~ \forall \hat{\mathbf{u}}(\cdot),\mathbf{u}(\cdot) \in \mathbb{U},\\ &(\hat{\mathbf{u}}(\tau)=\mathbf{u}(\tau) \textrm{a.e.} \tau \geq 0) \Rightarrow (\delta[\hat{\mathbf{u}}](\tau) = \delta[\mathbf{u}](\tau) \textrm{a.e.} \tau \geq 0)  \}.
    \end{aligned}
\end{equation}

The value of the differential game is as follows for a given $\bar{z} \in \mathcal{Z}$ is:
\begin{align}
    V(\bar{z}) := \inf_{\delta \in \Delta} \sup_{\bar{u} \in \mathbb{U}} \mathcal{V}(\bar{z},\bar{u}(\cdot),\delta[\bar{u}](\cdot)).
\end{align}
Reachable set $\mathcal{R}$ is the sub-zero level set of $V(\cdot)$: $\mathcal{R} = \{\bar{z} \in \mathcal{Z} ~ | ~ V(\bar{z}) \leq 0\}$ \cite{mitchell2005time}.   
    
\subsubsection{Hamilton-Jacobi Partial Differential Equation (PDE)} 
The value function can be solved as a solution of a PDE. To solve it as a PDE, $V(\cdot)$ is augmented with a time argument. $V(\cdot)$ is the value of the differential game for all time. We define $\bar{V}(z,t;T)$ as the value of the differential game from time $t \in (0,T)$ to terminal time $T \in [0,\infty)$. $\bar{V}(z,0;T)$ is equivalent to $V(z)$ as $T \rightarrow \infty$. 
In the principle of dynamic programming, $\bar{V}(z,0) := \bar{V}(z,0;T)$ can be attained by solving a modified  Hamilton-Jacobi-Issacs(HJI \cite{mitchell2005time}) PDE, with initial condition $\bar{V}(z,T) = l(z)$:
\begin{align}
        0 = \frac{\partial \bar{V}}{\partial t}(z,t)+\min \left\{0,\max_{u \in \mathbb{U}} \min_{d \in \delta[u]} \frac{\partial \bar{V}}{\partial z}(z,t)f(z,u,d)\right\}. \label{eq:HJIPDE}
\end{align}

In the context of safety analysis, if target set $\Omega$ is a set of unwanted states, for example, collision states, the reachable set $\mathcal{R}$ is a set of state that may end up collision, which is considered as unsafe. On the other hand, the complement $\mathcal{Z} \setminus \mathcal{R}$ is the set of states that there exists some input such that avoidance of $\mathcal{R}$ under all circumstances. In other words, safety can be guaranteed. Therefore, we can formulate safety analysis of the car following system as a reachability analysis.  

\subsection{Formulating safety analysis for FollowerStopper}



The two-car system can be described by a three dimensional system, where the components represent the relative distance between leading vehicle and the subject vehicle $x_{rel}(t) \in \mathbf{R}$, the relative speed between two vehicles  $v_{rel}(t)$, and absolute speed of the subject vehicle $v_{AV}(t) \in \mathbf{R}$. Let $z(t) = [x_{rel}(t),v_{rel}(t), v_{AV}(t)]^T.$ The dynamics of the two-car system is as follows:
\begin{align}
\dot{z}(t) 
&= \begin{bmatrix}
v_{rel}(t) \\
\lambda(v_{rel}(t)+v_{AV}(t))d(t) - \lambda(v_{AV}(t)) u(t) \\
\lambda(v_{AV}(t)) u(t)
\end{bmatrix} \nonumber \\  &:=  \bar{f}(z(t),u(t),d(t)),  \label{eq:3d_car_followeing_sys}
\end{align}
where $d(t) \in \mathbf{R}$ represents the lead vehicle acceleration; and $u(t) \in \mathbf{R}$ represents the subject vehicle acceleration, at time $t \in [0,\infty).$ $\lambda: \mathbf{R} \rightarrow \mathbf{R}$ is introduced to take into account speed constraints:
\begin{align}
    \lambda(y) = \begin{cases}
    1 &\mbox{if } y > 0 \\
    0 &\mbox{otherwise }
\end{cases}.
\end{align}
Note that $d(t)$ and $u(t)$ can be considered as inputs to the system, $d(t)$ being controlled adversarially by the known disturbance, and $u(t)$ being controlled by the FollowerStopper.

To synthesize the system's reachable sets, we formulate the problem as a min-max problem and we can generate the value function for FollowerStopper by solving the Hamiltonian in equation \eqref{eq:HJIPDE}. We formulate the min-max problem as follows:
\begin{align}
    &V^*(\bar{x}_{rel},\bar{v}_{rel}, \bar{v}_{AV})= \max_{u(\cdot)} \min_{d(\cdot)} \inf_{t \geq 0} (x_{rel}(t))\label{sys:3d_safety}\\
    \textrm{s.t.} \quad & \dot{z}(t) = \bar{f}(z(t),u(t),d(t)), \nonumber \\
    &u(t) = f_{AV}(x_{rel}(t),v_{rel}(t),v_{AV}(t)), \nonumber \\
    & d_{min}\leq d(t) \leq d_{max}, u_{min} \leq u(t) \leq u_{max}, \nonumber  \\
    & x_{rel}(0) = \bar{x}_{rel}, v_{rel}(0) = \bar{v}_{rel},
     v_{AV}(0) = \bar{v}_{AV}, \nonumber
\end{align}
where $f_{AV}:\mathbf{R} \times \mathbf{R} \times \mathbf{R} \rightarrow \mathbf{R}$ is an autonomous vehicle model consisted of controller dynamics and vehicle dynamics. Particularly, in this work, we are considering FollowerStopper introduced in section \ref{sec:FollowerStopper} and the vehicle model show in section \ref{sec:vehicleDynamicsModelA}.

\subsection{Vehicle dynamics model} \label{sec:vehicleDynamicsModelA}
Vehicle dynamics limits the speed response of the vehicle, which has impacts on the safety. Therefore, the dynamics of the vehicle should also be considered in the formulation of safety analysis. Given a $\tau \in \mathbf{R}_{>0}$, a first order ordinary differential equation (ODE) is considered as an approximation of dynamical response of the vehicle speed:
\begin{align}
\dot{v}_{AV}(t) = -\frac{1}{\tau} v_{AV}(t) + \frac{1}{\tau} v_{cmd}(t).
\label{eq:vehicle_dynamics_modelA}
\end{align}

where $v_{AV}(t)$ gives the speed of the autonomous vehicle at time $t$.

\subsection{Driving data and safe set}
Real world driving data was collected with on-board sensors of production vehicles. More than $1100$ miles of driving data has been gathered. We collected data including relative distance, relative speed, subject vehicle speed, and subject vehicle acceleration. Driving data are used to estimate acceleration/deceleration bounds for calculation of safe set and to evaluate the generated safe set. By overlaying collected driving data and the safe set, we can assess whether FollowerStopper can safely handle the driving condition experienced by a human driver. Furthermore, car-following behaviors of FollowerStopper can be compared to human drivers by running simulations with collected data.

\section{{Using reachability for safety analysis of FollowerStopper}} \label{sec:safetyAnalysis4FollowerStopper}



\subsection{Safe set for FollowerStopper}
\begin{figure}[tpb]
    \vspace{+0.6em}
    \centering
    \includegraphics[trim={0.0cm 0.0cm 0.0cm 0.0cm},clip,width=0.7\linewidth]{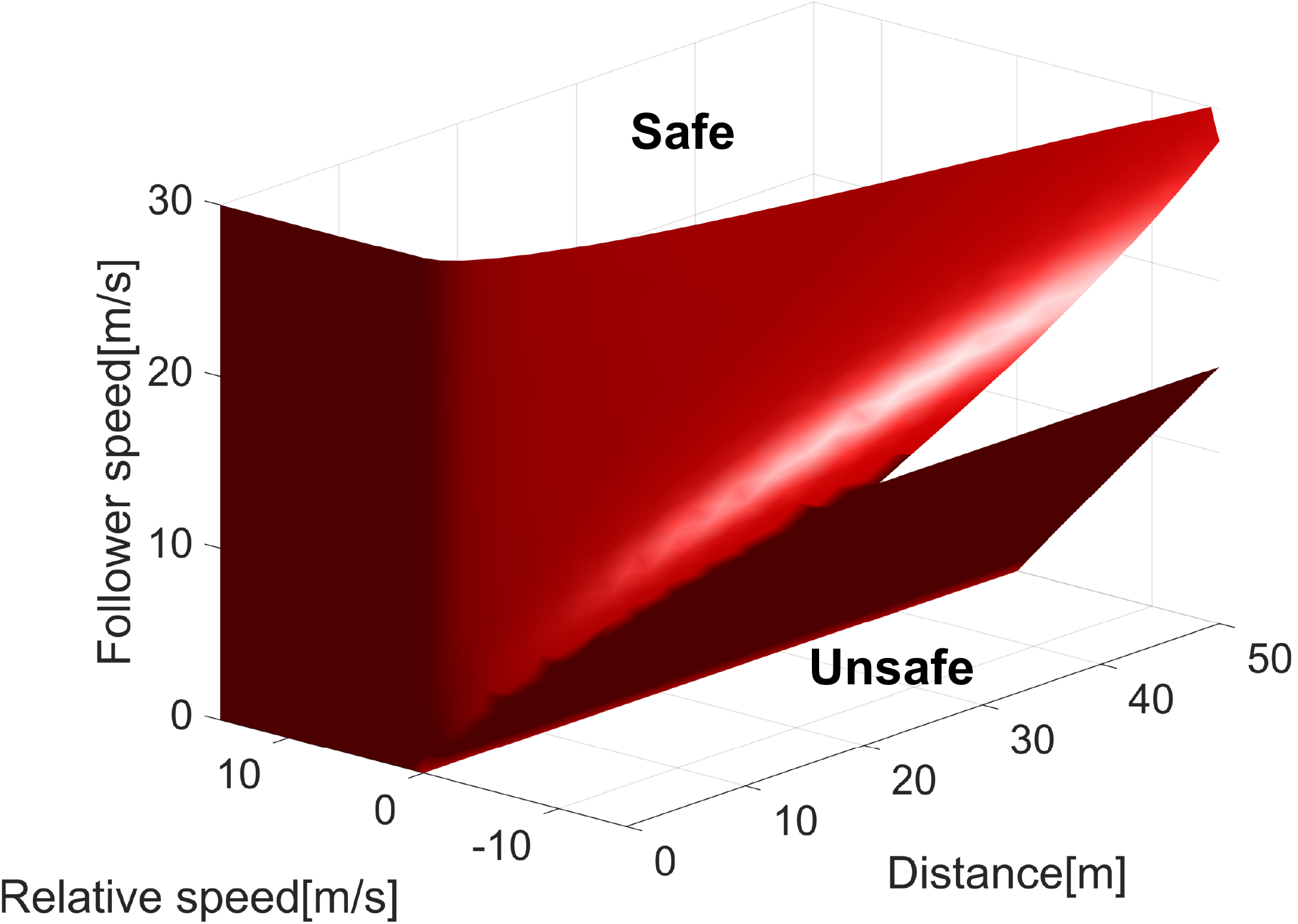}
    \caption{Safe set for FollowerStopper. The red surface is the boundary of the safe set. From this angle, the safe set is on the other side of the surface.}
    \label{fig:3d_safety_set}
\end{figure}

The safe set for FollowerStopper is shown below in \mbox{figure \ref{fig:3d_safety_set}}. The red surface renders the boundary of the safe set in the space where the distance is between $0$ and $50$m, the relative speed is between $-15$m/s and $15$m/s, and the follower speed is between $0$ and $30$m/s. The safe set is the region enclosed by the surface. For clarity, we show the projections of the safe set at different follower speeds in \mbox{figure \ref{fig:FS_modelA_SafetySet_fieldTestData}.} Yellow curves are boundaries of the safe set for all graphs. The safe sets are the regions above the curves. 

\subsection{FollowerStopper safe set and driving data}
\label{sec:drivingDataAndSafety}
\begin{figure}[htpb]
    \centering
    \includegraphics[clip,width=0.95\linewidth]{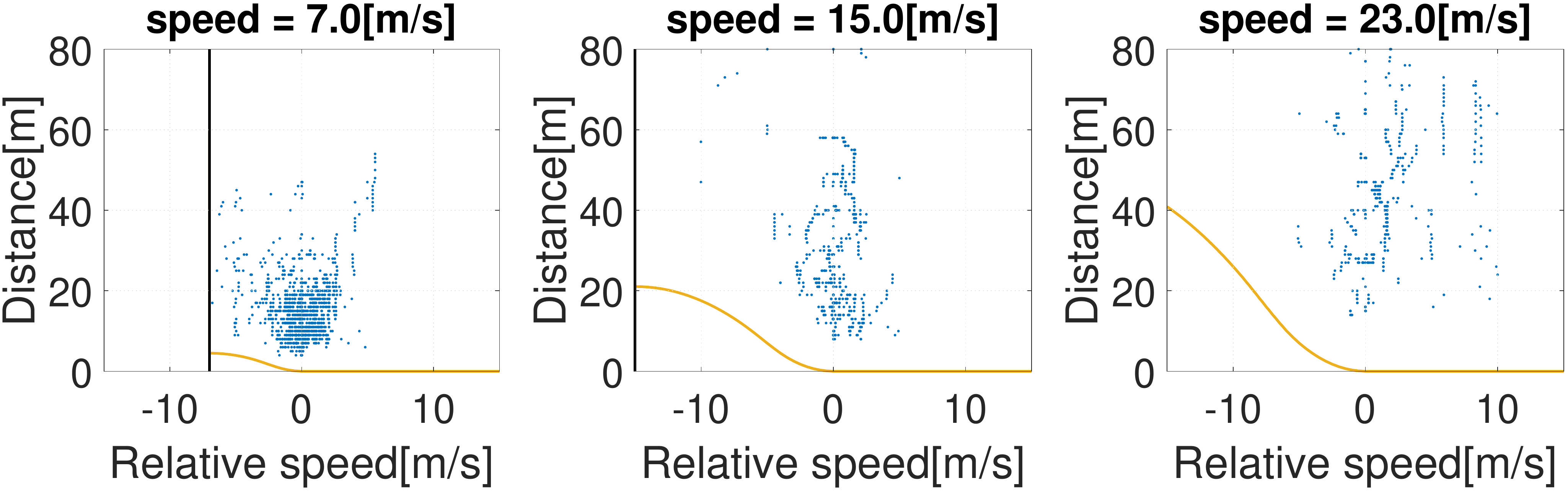}
    \caption{Safe set of FollowerStopper projections of safe sets on 2D plane along with filed test data at different speeds. The yellow curve is the safety boundary of FollowerStopper (equation \eqref{sys:3d_safety}). Regions above the curves are provably safe. }
    \vspace{-0.em}
    \label{fig:FS_modelA_SafetySet_fieldTestData}
\end{figure}

We use human driving data to assess the safe set of FollowerStopper. We observed that all human driving data points we collected so far are inside the safe set. For clarity, fractions of field test data are shown with projections of safe sets at a few different speeds in Figure \ref{fig:FS_modelA_SafetySet_fieldTestData}. Given that the safety of any state in the safe set can be guaranteed by FollowerStopper, theoretically, all driving data experienced by human driver can be handled by FollowerStopper with safety guarantee.



\subsection{Simulation with driving data}
\begin{figure}
    \vspace{+0.5em}
    \begin{subfigure}{0.23\textwidth}
    \centering
     \includegraphics[trim={0.0cm 0.0cm 0.0cm 0.0cm},clip,width=\textwidth]{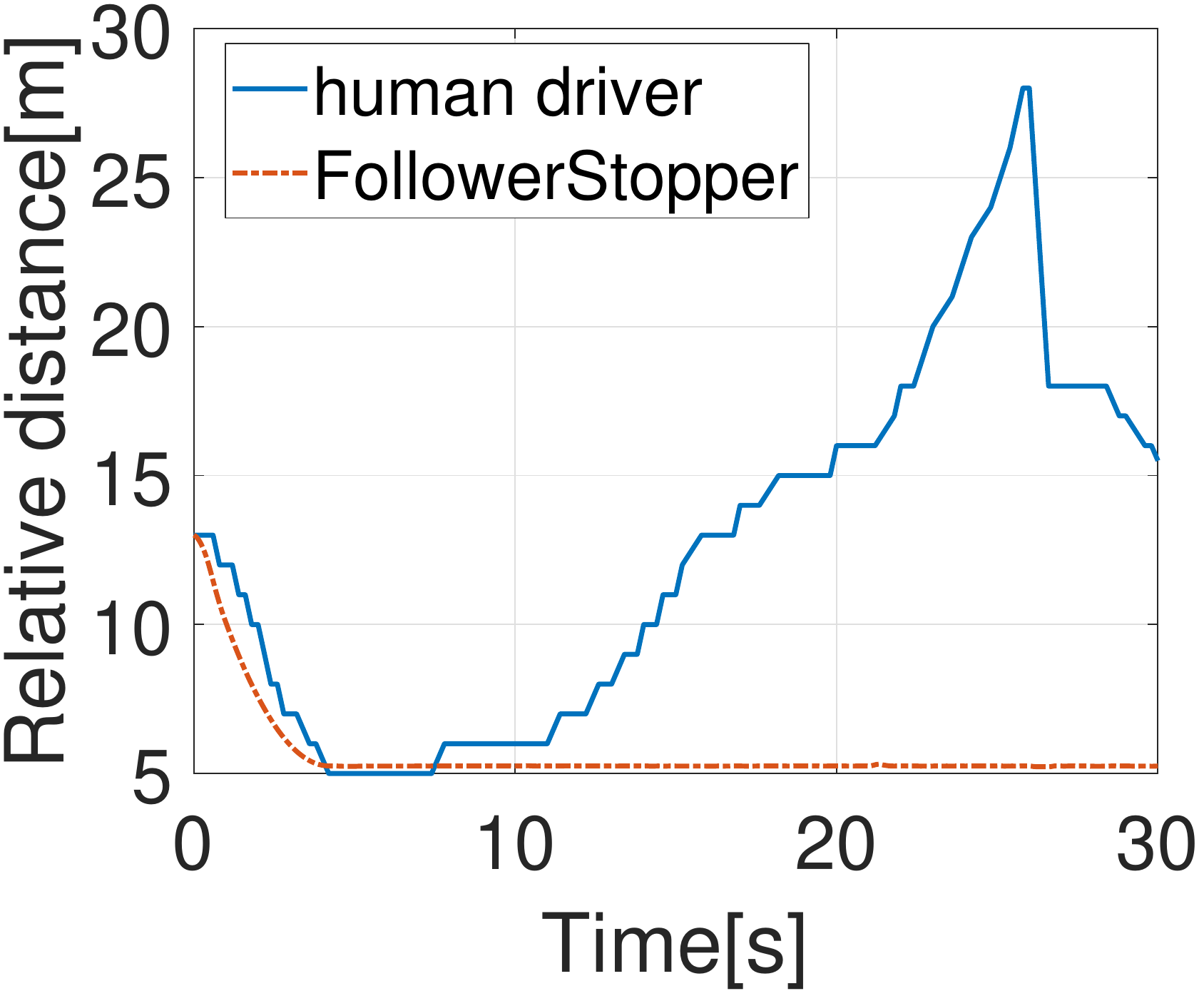}
        \label{fig:FollowerStopper_vs_human_dist_low_modelA}
     \end{subfigure}
     \begin{subfigure}{0.23\textwidth}
         \centering
         \includegraphics[trim={0.0cm 0.0cm 0.0cm 0.0cm},clip, width=\textwidth]{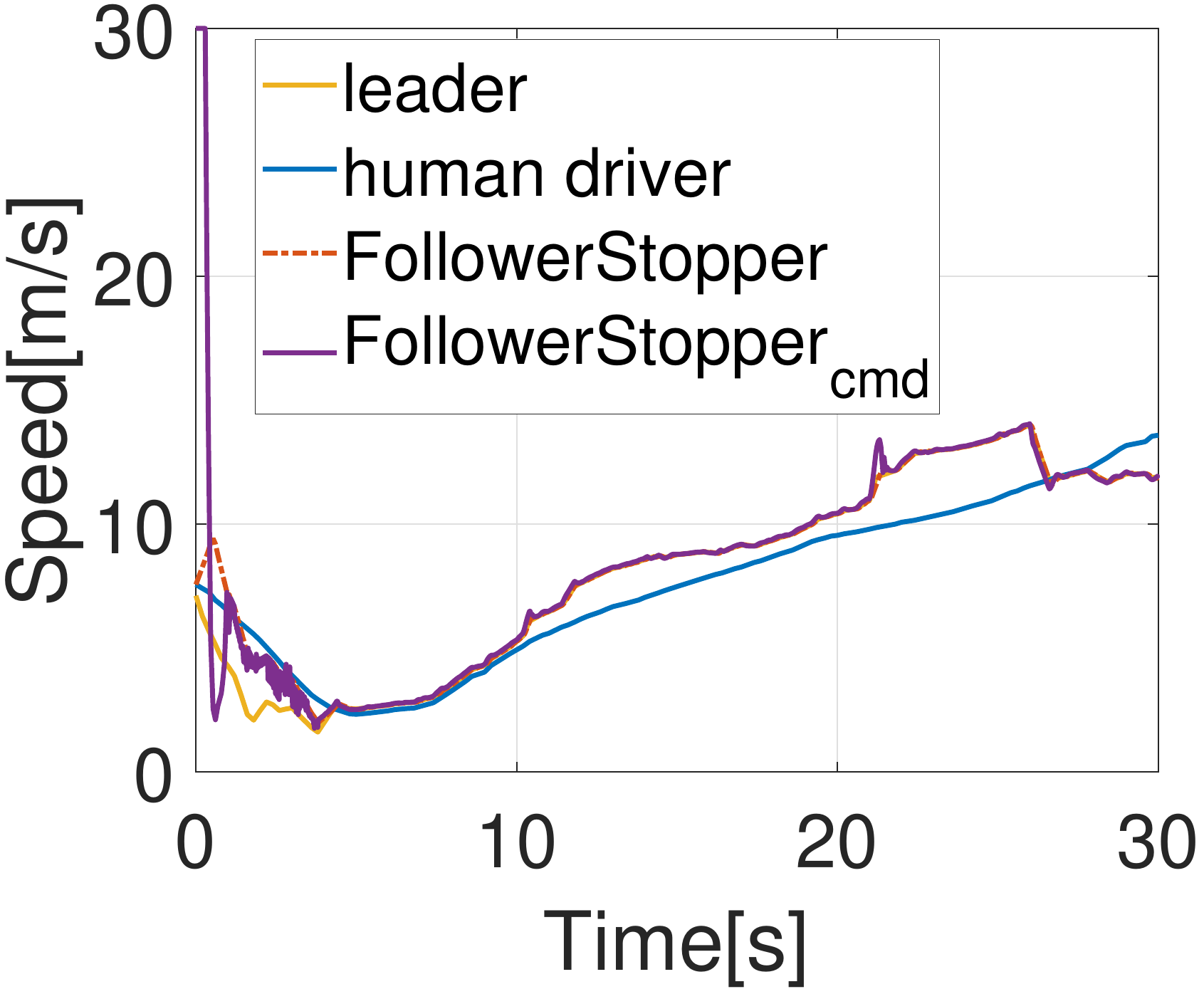}
         \label{fig:FollowerStopper_vs_human_speed_low_modelA}
     \end{subfigure}    
     \vspace{-1.5em}
     \caption{ Comparison between FollowerStopper and human driver at low speed.  \textbf{Left}: Relative position. \textbf{Right}: Speed.  In left, the blue curve shows the relative distance experienced by human driving in the real world, and the red curve is simulated distance for FollowerStopper following the same leading vehicle. In right, the yellow curve represents the measured speed of the leading vehicle and the blue curve represents the measured speed of the subject vehicle in the real world. The red curve and the purple curve respectively represent the speed, and the speed command of FollowerStopper, in simulation. Human driver increases inter-vehicle space as driving speed increases. FollowerStopper follows the leading vehicle closely at around 5 meters.}
     \label{fig:FollowerStopper_vs_human_low_modelA}
\end{figure}

\begin{figure}
     \centering
     \begin{subfigure}{0.23\textwidth}
         \centering
     \includegraphics[trim={0.0cm 0cm 0.0cm 0cm},clip,width=\textwidth]{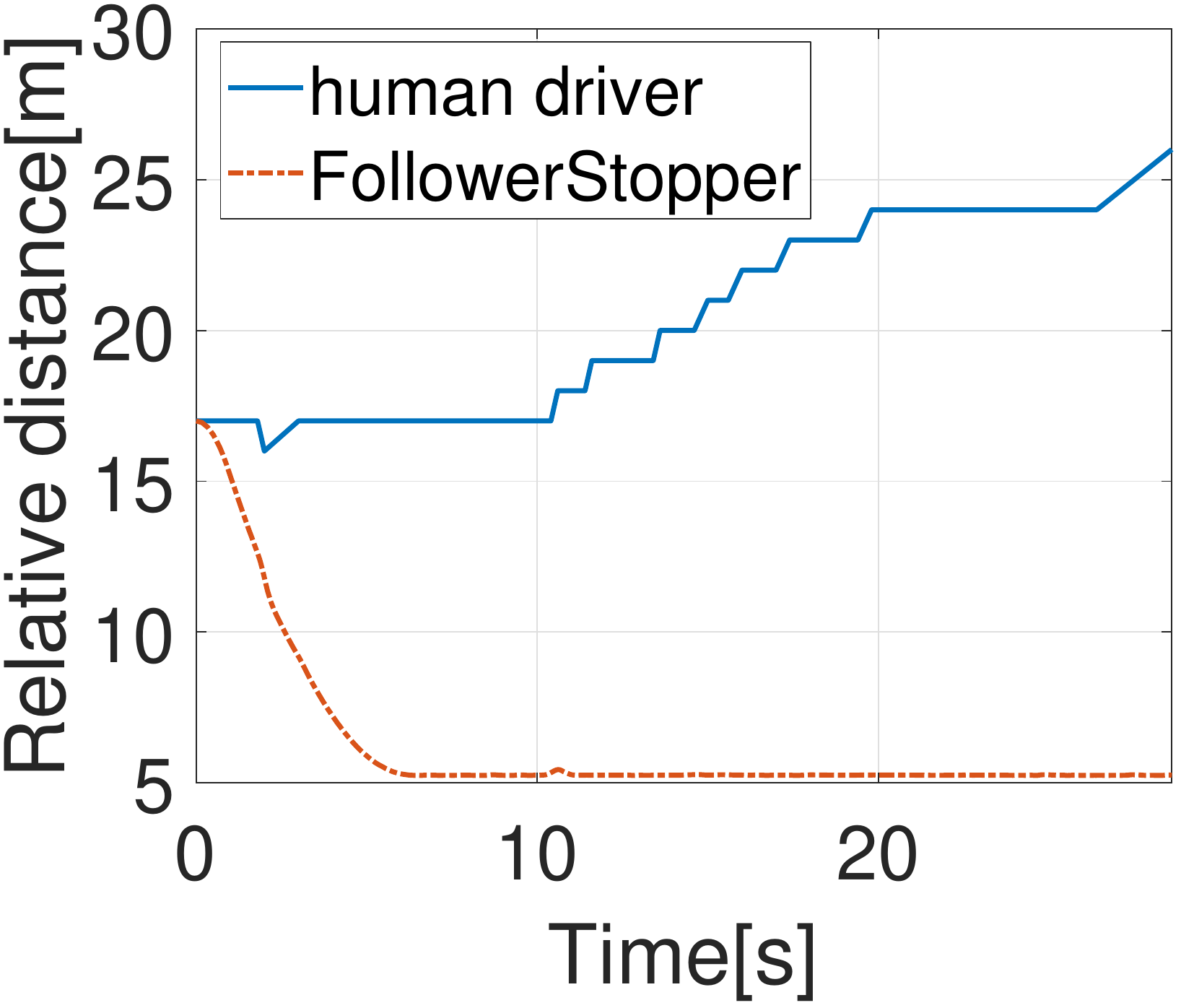}
         \label{fig:FollowerStopper_vs_human_dist_high_modelA}
     \end{subfigure}
     \begin{subfigure}{0.23\textwidth}
         \centering
         \includegraphics[trim={0cm 0cm 0.0cm 0cm},clip,width=\textwidth]{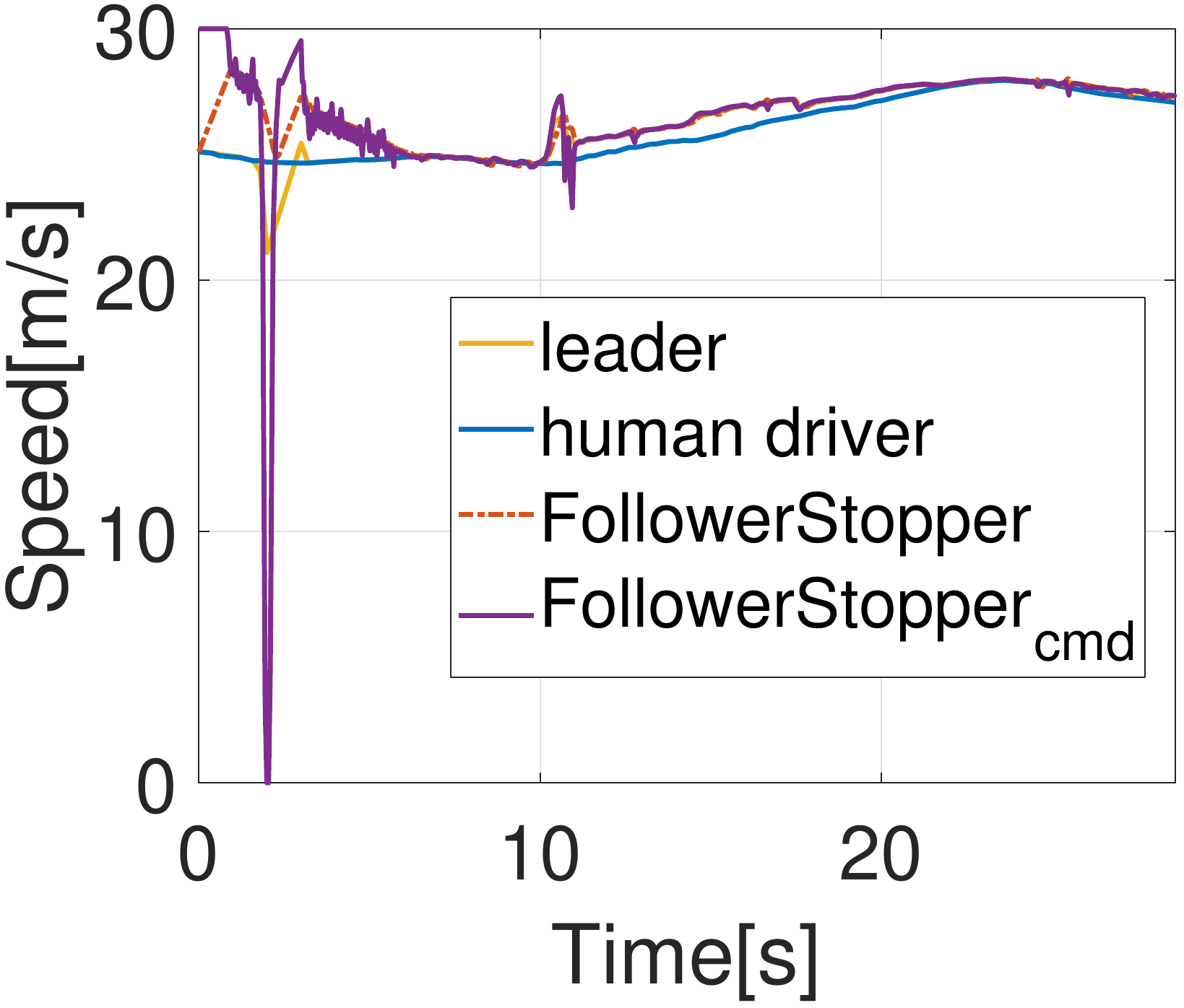}
         \label{fig:FollowerStopper_vs_human_speed_high_modelA}
     \end{subfigure}
     \vspace{-1.5em}
     \caption{Comparison between FollowerStopper and human driver at high speed. \textbf{Left}: Relative position. \textbf{Right}: Speed. FollowerStopper follows the leading vehicle closely with a short gap, while human driver keeps a larger gap at high speed.}
     \vspace{-1.5em}
     \label{fig:FollowerStopper_vs_human_high_modelA}
\end{figure}

We compare human driving behavior against FollowerStopper in simulation. We simulate the FollowerStopper following human driving leading vehicles. We show two instances: figure \ref{fig:FollowerStopper_vs_human_low_modelA} for low speed scenario and \mbox{figure \ref{fig:FollowerStopper_vs_human_high_modelA}} for high speed scenario. At both low-speed and high-speed scenarios, simulation results show that FollowerStopper follows the leading vehicle without collision; however, the following distance is short at high speed. Even though it is considered safe based on the distance-based safety criterion, it is not desirable, because short distance at high speed allows subject vehicle less time to respond to unexpected incidents. Therefore, we modified safety analysis considering time headway to take into account speed in the safety criterion.


\section{Safety analysis with time headway} \label{sec:safetSetWithTimeHeadway}
We add time headway to the original formulation of the safety analysis to take into account speed in the safe following distance. Particularly, we change the fist line of equation \eqref{sys:3d_safety} as follows and keeps the rest of the constraints the same.
\begin{align}
    V^*(\bar{x}_{rel},\bar{v}_{rel}, \bar{v}_{AV})&= \max_{u} \min_d \inf_{t \geq 0} (x_{rel}(t)-h \cdot v_{AV}(t)),  \label{sys:3d_safety_with_time_headway}
\end{align}
where $h \in \mathbf{R}$ is a design parameter. In the equation \eqref{sys:3d_safety_with_time_headway}, the safe distance is now defined based on the time headway. It is considered as unsafe if the time headway is shorter than $h$. We pick $h = 0.4$, which is the minimum time headway shown in the driving data. 

\section{Modified FollowerStopper} 
\label{sec:Modify_FollowerStopper}
We use the reformulated safety verification to investigate FollowerStopper. Results shows that FollowerStopper is not safe based on the time headway-based safety criterion. The FollowerStopper is improved by changing adaptation regions to take into account time headway.
\subsection{Formulation of the modified FollowerStopper}
\begin{figure}[htpb]
\begin{subfigure}{0.81\textwidth}
    \includegraphics[width=0.3\linewidth]{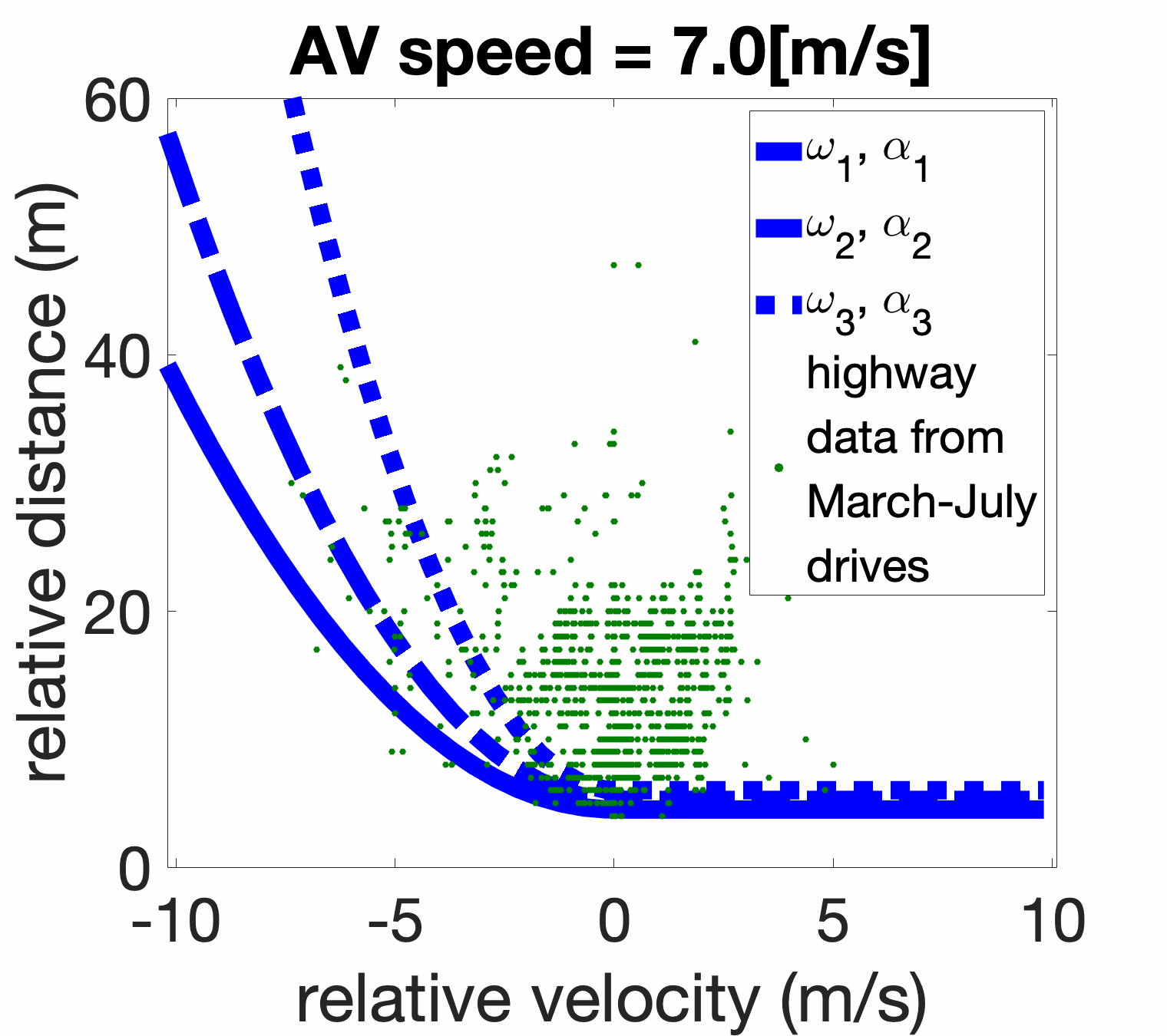}
    \includegraphics[width=0.3\linewidth]{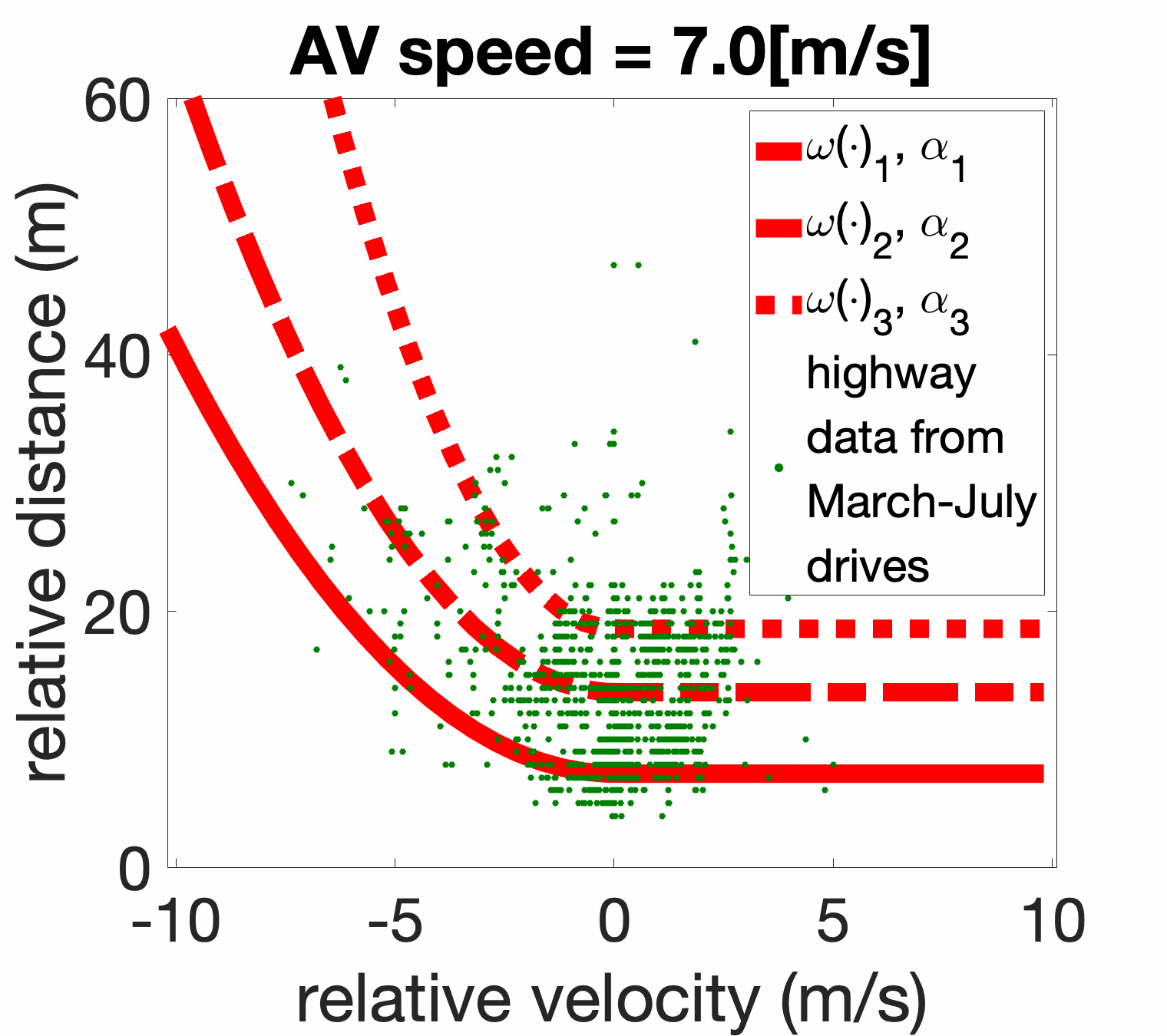}
\end{subfigure}

\begin{subfigure}{0.81\textwidth}
    \includegraphics[width=0.3\linewidth]{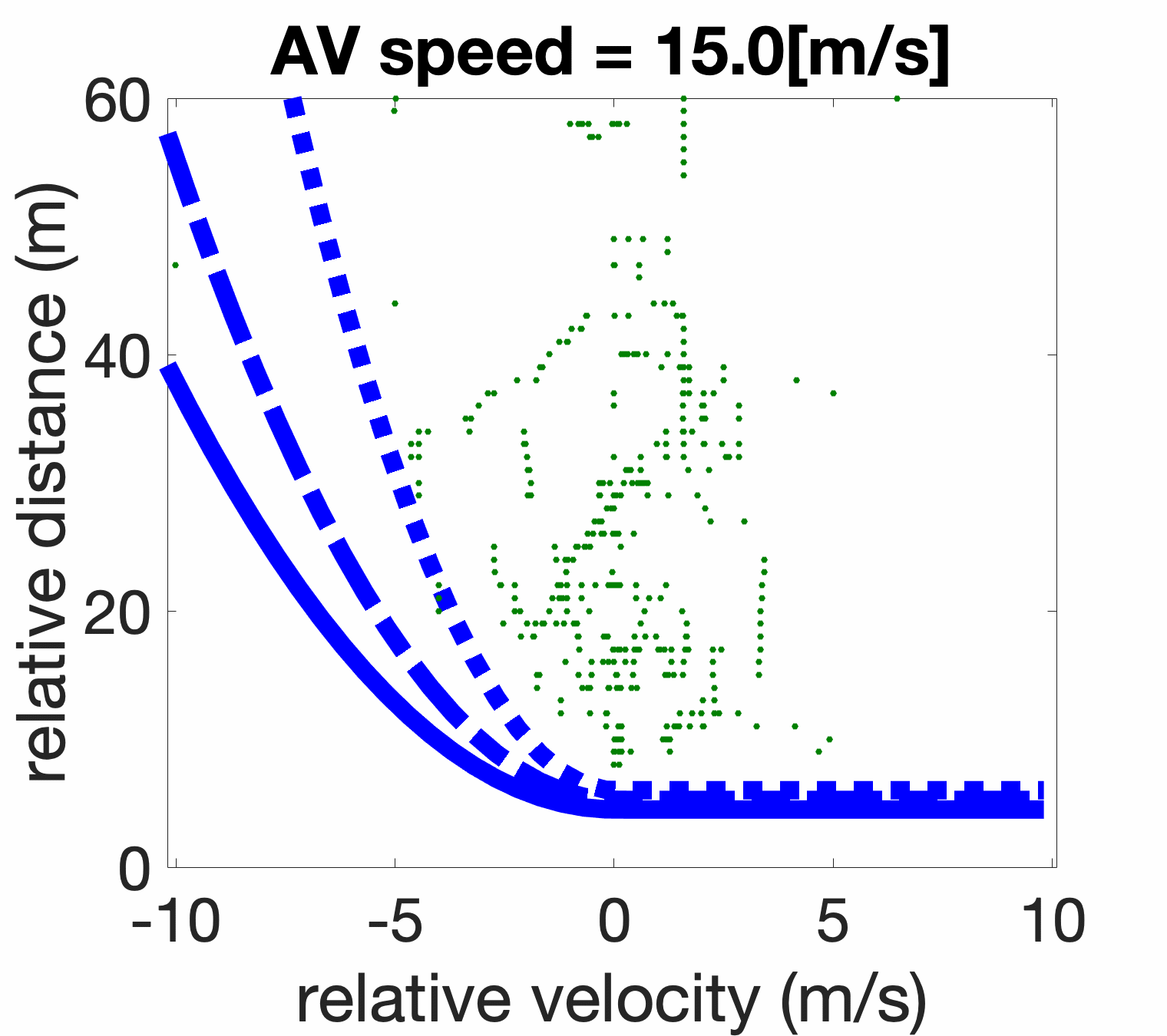}
    \includegraphics[width=0.3\linewidth]{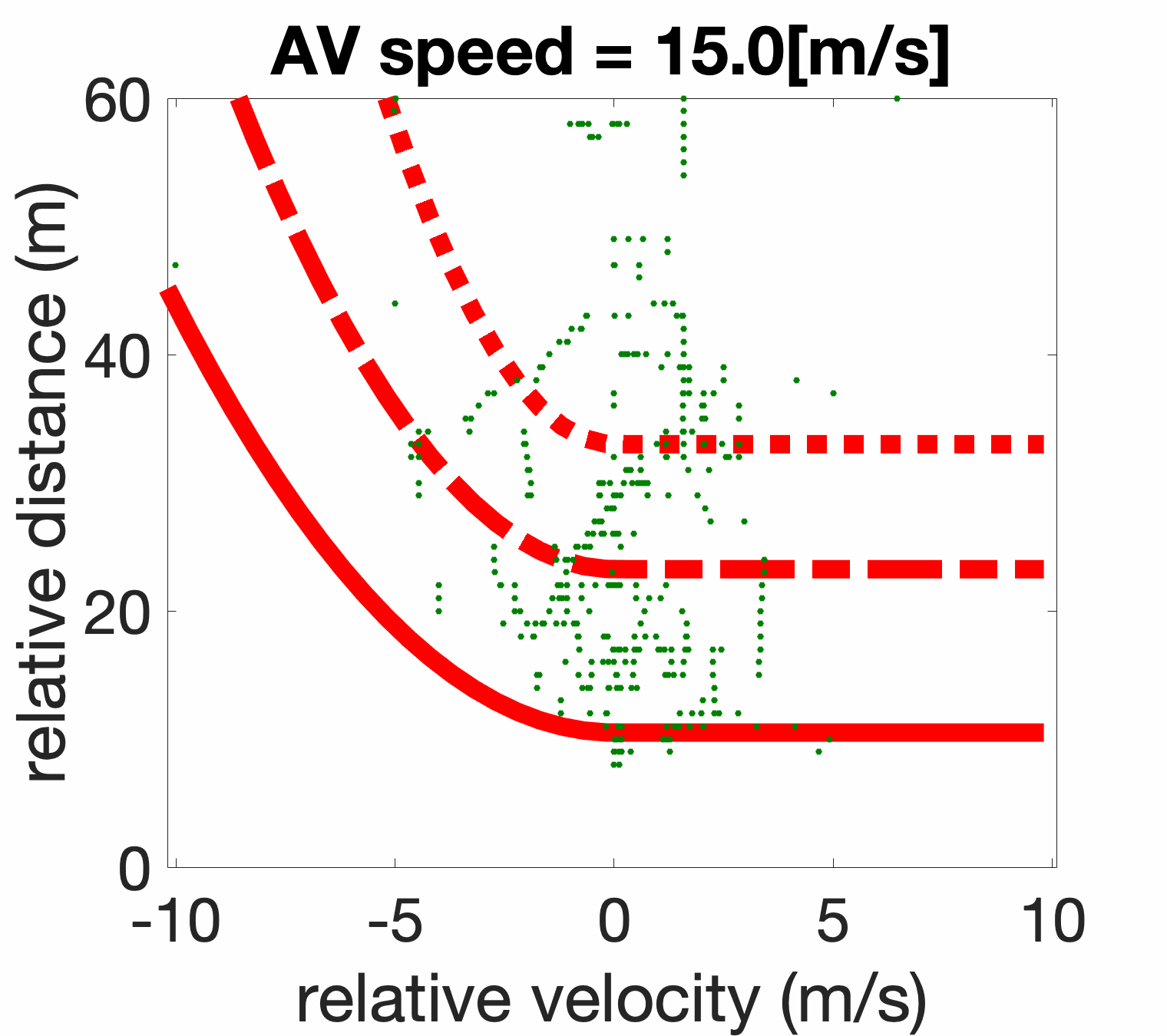}
\end{subfigure}

\begin{subfigure}{0.81\textwidth}
    \includegraphics[width=0.3\linewidth]{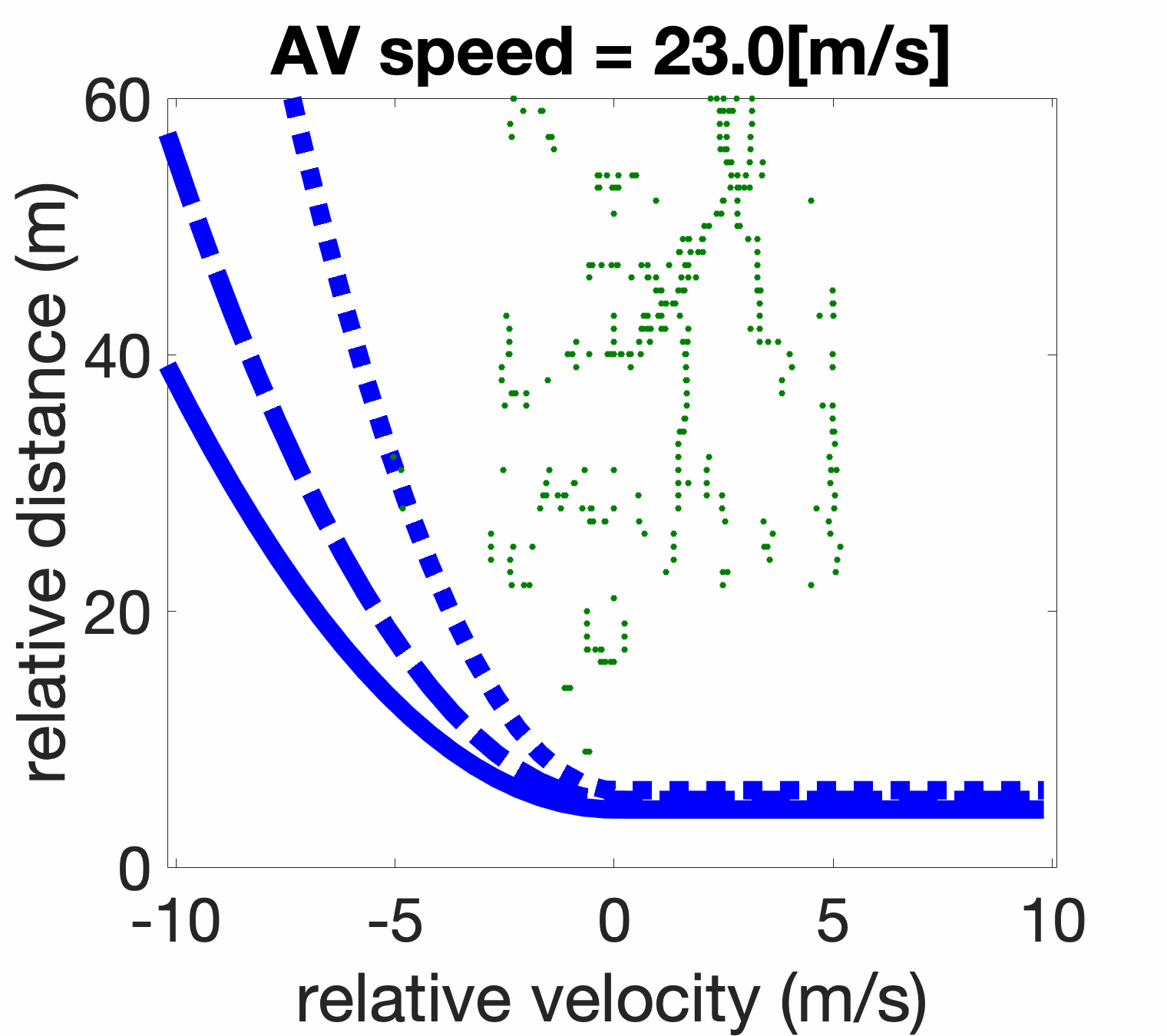}
    \includegraphics[width=0.3\linewidth]{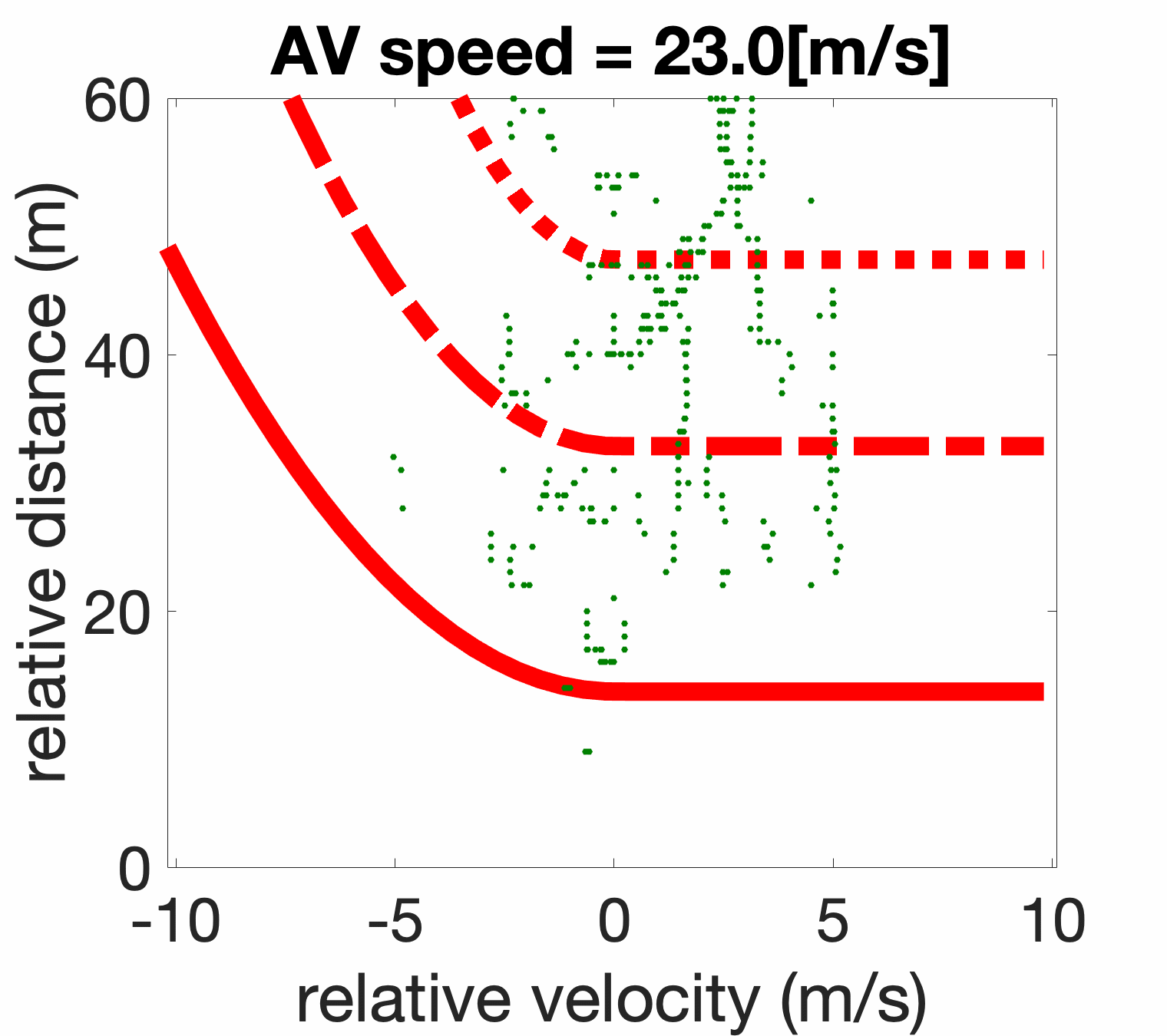}
\end{subfigure}
\caption{Modified FollowerStopper - comparison between old quadratic curve with various switching regions from Figure  \ref{fig:quadratic_curve} (represented in blue) and switching regions defined from equation (\ref{Eq:new_FS_regions}) (represented in red). The new regions defined by equation (\ref{Eq:new_FS_regions}) follows the data with changes in speed - intuitively this tell us that the $\omega_j$ in our design parameters should be velocity varying.} 
\label{fig:compare_new_old_plots}
\end{figure}

In order to satisfy the time headway-based criterion of safety, adaptation regions of FollowerStopper \eqref{eq:FS_segmentation} are modified with time headway $h_j \in \mathbf{R}$ as follows: 
\begin{equation}
    \bar{x}_j = \omega_j + \frac{1}{2\alpha_j} {v_{rel}^*}^2 + h_j \cdot v_{AV}, \quad j = 1,2,3. 
    \label{Eq:new_FS_regions}
\end{equation}
Here, we pick $h_1 = 0.4$, $h_2 = 1.2$, and $h_3 = 1.8$. Figure \ref{fig:compare_new_old_plots} shows the comparison between the two switching designs.

\subsection{Driving data and safety}
\begin{figure}[htpb]
    \centering
\includegraphics[clip,width=0.95\linewidth]{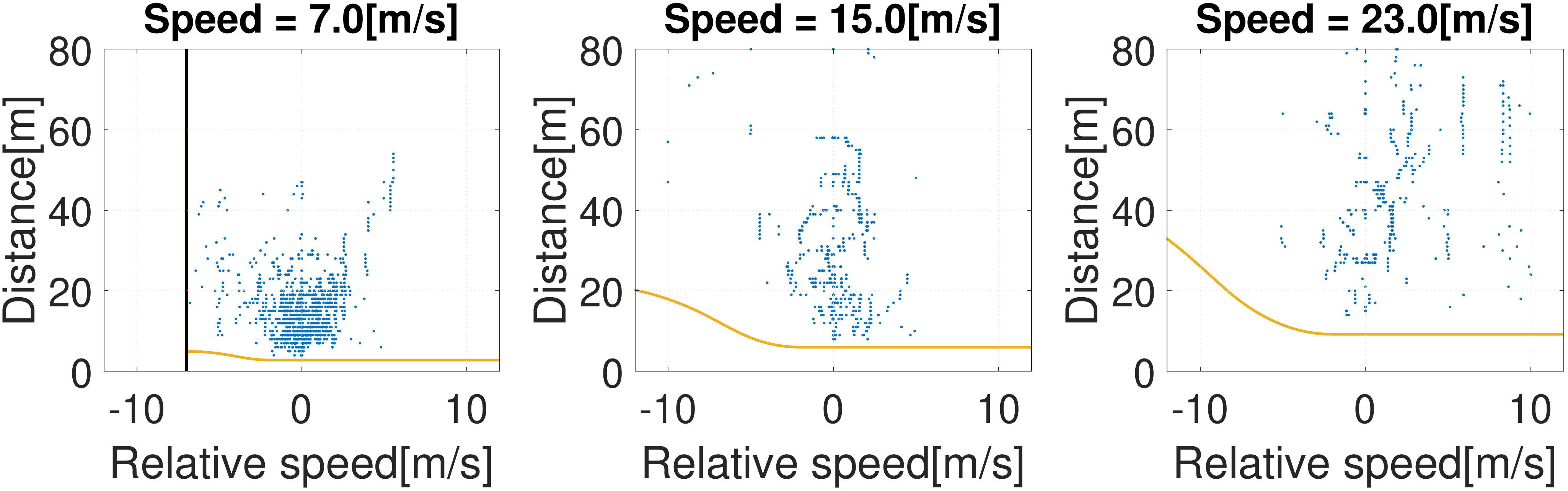}
    \caption{Time headway-based safe set of the modified FollowerStopper - projections of safe sets on 2D plane along with field test data at different speeds. Yellow curves are the safety boundaries of the modified FollowerStopper. Regions above the curves are considered safe. }
    \vspace{-0.4em}
    \label{fig:RobustFS_NewSafetySet_vs_Data_HJI_toolbox}
\end{figure}

Figure \ref{fig:RobustFS_NewSafetySet_vs_Data_HJI_toolbox} shows the time headway-based safe set of the modified FollowerStopper at different speeds. The major difference compared to figure \ref{fig:FS_modelA_SafetySet_fieldTestData} is that the minimum safe distance is not zero but increased with vehicle speed. Given the safe set of the modified FollowerStopper still covers all driving data, the modified FollowerStopper is guaranteed to have minimum time headway to the leading vehicle for all time.
    
\subsection{Simulation with driving data}
\begin{figure}[htpb]
\centering
\begin{subfigure}{0.23\textwidth}
\centering
    \includegraphics[width=\textwidth]{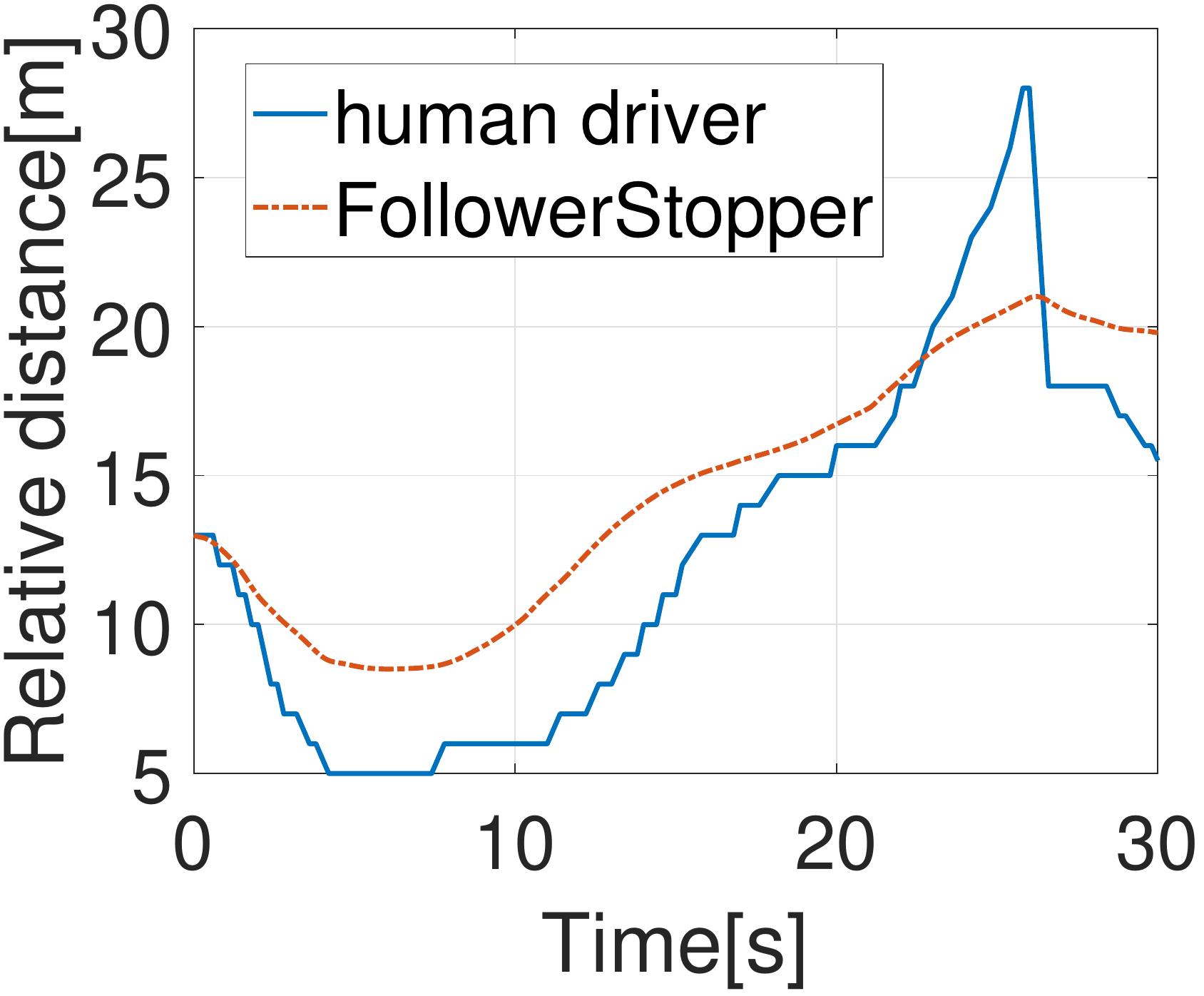}
    \label{fig:ModifiedFollowerStopper_vs_human_dist_low_modelA}
\end{subfigure}
\begin{subfigure}{0.23\textwidth}
    \includegraphics[width=\textwidth]{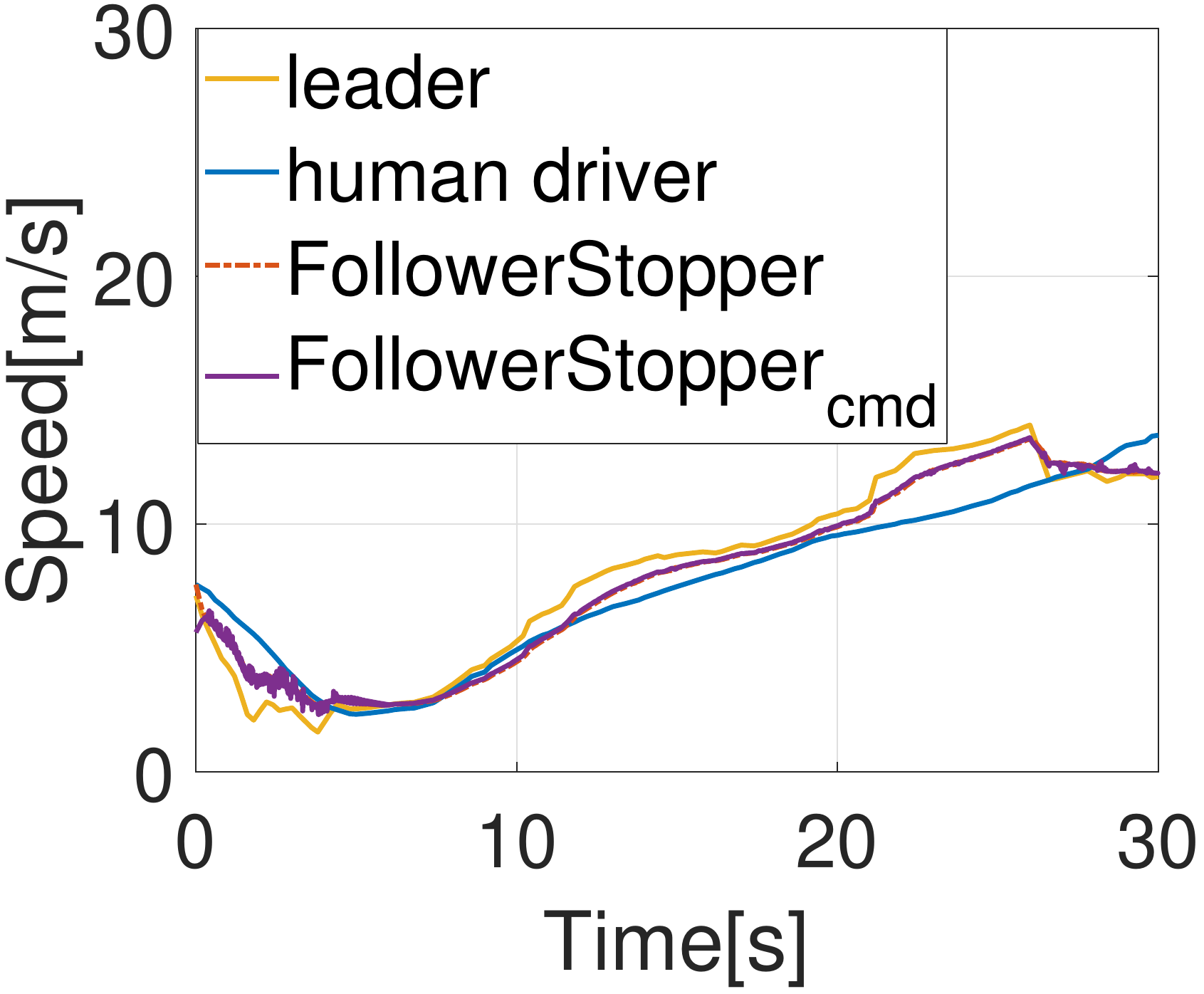}
    \label{fig:ModifiedFollowerStopper_vs_human_speed_low_modelA}
    \end{subfigure}
    \vspace{-1.5em}
    \caption{Comparison between modified FollowerStopper and human driver at low speed. \textbf{Left}: Relative position. \textbf{Right}: Speed. The modified FollowerStopper increases inter-vehicle space at higher speed, which is similar to human driver. }
    \vspace{-1.0em}
    \label{fig:ModifiedFollowerStopper_vs_human_low_modelA}
\end{figure}

\begin{figure}[htpb]
\centering
\begin{subfigure}{0.23\textwidth}
\centering
    \includegraphics[width=\textwidth]{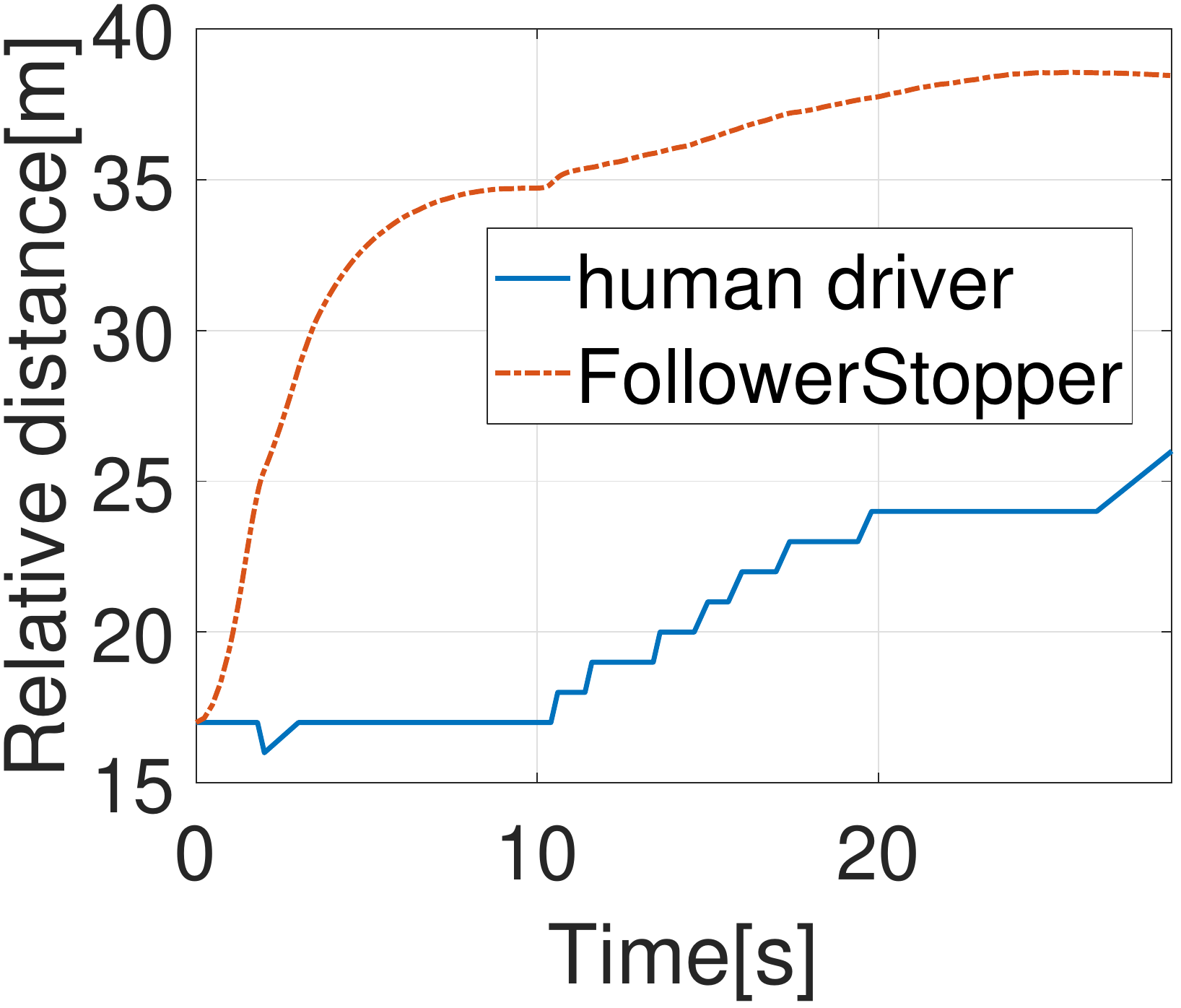}
    \label{fig:ModifiedFollowerStopper_vs_human_dist_high_modelA}
\end{subfigure}
\begin{subfigure}{0.23\textwidth}
    \includegraphics[width=\textwidth]{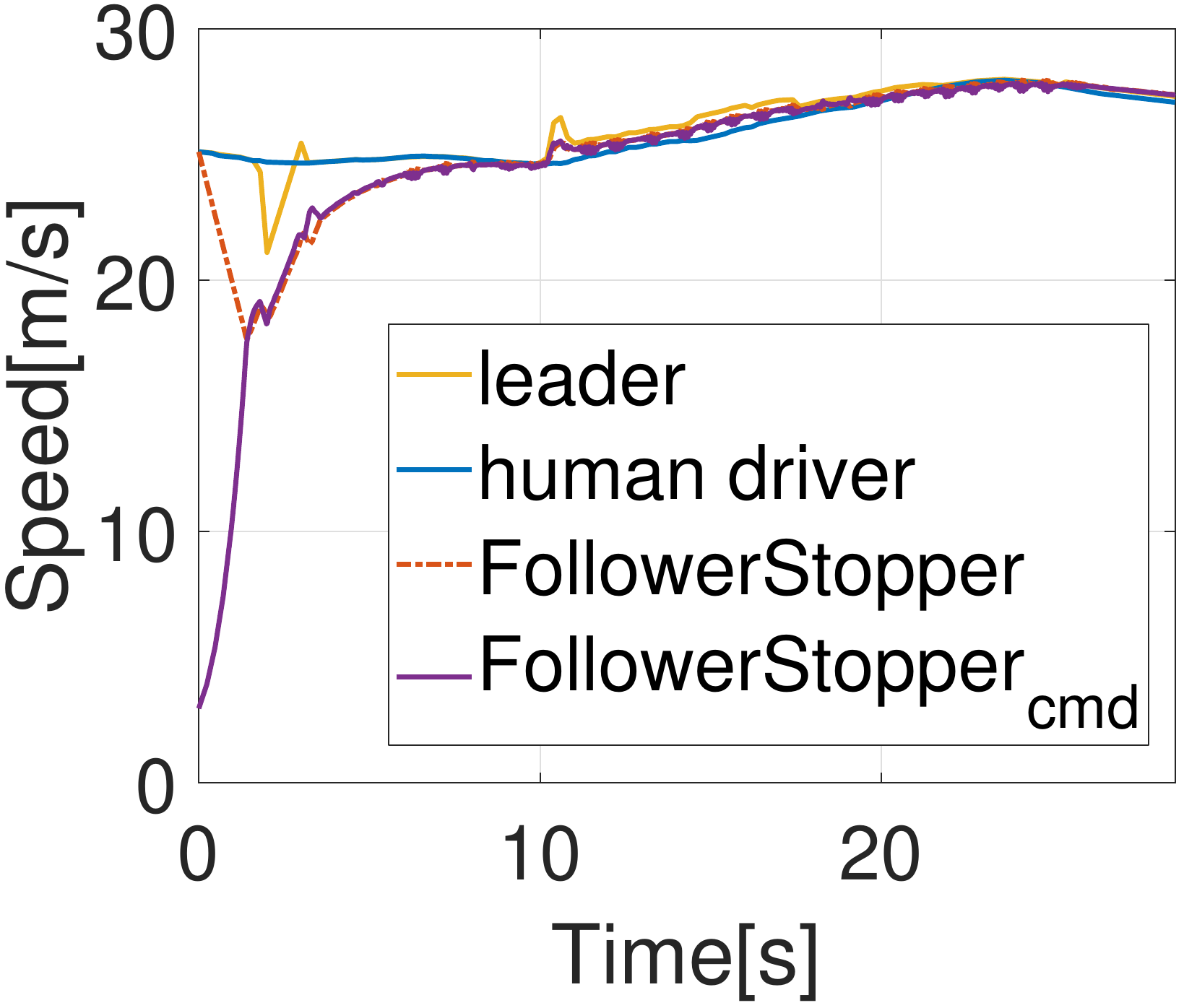}
    \label{fig:ModifiedFollowerStopper_vs_human_speed_high_modelA}
    \end{subfigure}
    \vspace{-1.5em}
    \caption{Comparison between modified FollowerStopper and human driver at high speed. \textbf{Left}: Relative position. \textbf{Right}: Speed.  Comparing to figure \ref{fig:FollowerStopper_vs_human_low_modelA}, the modified FollowerStopper is not driving at closely to the leading vehicle, but increasing inter-vehicle space at high speed.}
    \label{fig:ModifiedFollowerStopper_vs_human_high_modelA}
\end{figure}

Similarly, we simulate the modified FollowerStopper with human driving data to validate the response of the modified FollowerStopper. Simulation results with human driving data are shown in figure \ref{fig:ModifiedFollowerStopper_vs_human_low_modelA} and figure \ref{fig:ModifiedFollowerStopper_vs_human_high_modelA}. Instead of following the leading vehicle at a short gap, the modified FollowerStopper following distance is contingent on vehicle speed, the larger distance at higher driving speed, which is more like human driver's following distance.

\section{Conclusions} \label{sec:conclusions}
We verify the safety of the FollowerStopper with reachability analysis. We firstly use a distance-based safety criterion to verify FollowerStopper by showing that there is a set of states that safety can be guaranteed. However, by running a simulation with real-world driving data, we find that FollowerStopper is following the leading vehicle closely even at high speed, which is undesirable. A more stringent safety criterion is needed to avoid considering such behavior as safe. We incorporate time headway in the safety analysis. FollowerStopper is not considered safe anymore under the new safety analysis. We modified the FollowerStopper to meet the new safety criterion. All human driving data is still covered by the new safe set, which means modified FollowerStopper is theoretically as safe as a human driver. Simulation results with human-driving data validated the modified FollowerStopper. In the future, the analysis will be extended with more detailed vehicle dynamics models considering physical limitations of real systems and the new design will be validated on vehicles in the real world. 



{\footnotesize
\section*{ACKNOWLEDGMENT}
The authors would like to thank Jonathan Lee for providing valuable comments for this work. This material is based upon work supported by the National Science Foundation under Grant Numbers CNS-1837244 (A. Bayen), CNS-1837652 (D. Work), CNS-1837481 (B. Piccoli), CNS-1837210 (G. Pappas), CNS-1446715 (B. Piccoli), CNS-1446690 (B. Seibold), CNS-1446435 (J. Sprinkle, R. Lysecky), CNS-1446702 (D. Work). Any opinions, findings, and conclusions or recommendations expressed in this material are those of the author(s) and do not necessarily reflect the views of the National Science Foundation.
This material is based upon work supported by the U.S. Department of Energy’s Office of Energy Efficiency and Renewable Energy (EERE) award number CID DE-EE0008872. The views expressed herein do not necessarily represent the views of the U.S. Department of Energy or the United States Government.}


\bibliographystyle{IEEEtran}
\bibliography{main-bibliography}





\end{document}